\documentclass[aps,prd,groupedaddress,showpacs]{revtex4}
\usepackage{graphicx,epsf,amssymb,amsmath,latexsym}

\newcommand{\be}{\begin{eqnarray}}
\newcommand{\ee}{\end{eqnarray}}
\renewcommand{\d}{\partial}

\numberwithin{equation}{section}

\begin{document}

\begin{flushright}
SISSA 90/2007/A\\DAMTP-2007-117
\end{flushright}

%
\title{On interactions of higher spin fields with gravity and branes in $AdS_5$}

\vspace{.3in}

\author{Cristiano Germani}
\email{germani@sissa.it}
\affiliation{SISSA and INFN, via Beirut 4, 34014 Trieste, Italy}
\author{Arabella Schelpe}
\email{C.A.C.Schelpe@damtp.cam.ac.uk}
\affiliation{D.A.M.T.P., Centre for Mathematical Sciences, University of Cambridge,\\
Wilberforce road, Cambridge CB3 0WA, England}
\vskip.3in
\begin{abstract}
We construct actions of
higher spin fields interacting with gravity
on $AdS_5$ backgrounds such that the Compton scattering amplitudes of
the interaction are tree-level unitary.
We then consider higher-spin
fields in the Randall-Sundrum scenario. There, in the fermionic case,
we construct a tree-level unitary action of higher spin
fields interacting with branes and linearised gravity.
In the bosonic case we show that this is not in general possible. A tree-level unitary action of
bosonic higher spins interacting with linearised gravity and branes is only
possible in the following cases: The brane is a pure tension brane and/or Dirichlet
boundary conditions are imposed thereby making bosonic
higher spin fields invisible to a brane observer. We finally show that higher spins in Randall-Sundrum II braneworlds
can only be produced by (decay into) gravitons at trans-Planckian
scales. We end by commenting on the possible relevance of higher-spin
unparticles as
Dark Matter candidates.

\end{abstract}
\maketitle
%
\section{Introduction}

The problem of consistent higher-spin (HS) gauge theories is a
fundamental problem in field theory. After the description of
their free dynamics~\cite{Fronsdal:1978rb},\cite{Fang:1978wz},
only negative results for their interactions were
obtained~\cite{Aragone:1979hx},\cite{NWE} (see however \cite{ouvry0}). For example, it was realised
that HS fields cannot consistently minimally interact with gravitons and/or with Standard Model fields (SM) \cite{WW} in
a flat background.
However, in the case of an Anti-deSitter ($AdS$) gravitational background, by allowing additional non-minimal gauging,
one may introduce
counter terms, which make the interaction of HS fields
with gravitons well-defined. By appropriate completion of the interactions,
Vasiliev equations can be found
\cite{Vass},\cite{Vasss},\cite{Sezgin}, which are the generally covariant field equations for massless
HS gauge fields describing their consistent interaction with
gravitons.
In this framework HS interactions have been previously
discussed \cite{alk1, alk2, alk3, alk4}, however phenomenological
predictions are much harder to extract than in the framework we adopt
in this paper.

Given this theoretical basis, it is natural to wonder whether HS
particles can in principle exist in Nature as a four-dimensional
$AdS$ background seems to be incompatible with observations. However, in the past few years, much research
has been focused on so-called `braneworld' models, pioneered by \cite{Randall:1999ee}. In \cite{Randall:1999ee},
our Universe is
a four-dimensional hypersurface, a brane, embedded in an $AdS_5$ spacetime (bulk).
The Standard Model of particle physics (SM) is then supposed to live on the brane. In particular, extensions
of this model
produce interesting cosmological backgrounds (see for example \cite{langlois},\cite{KK},\cite{turok},\cite{sling}).
In this model,
HS fields can naturally live in the
extra-dimensional space.

In this paper we look at both the interactions of HS with gravitons in
pure $AdS_5$ and in braneworlds. Unlike
\cite{Vass, Vasss, Sezgin, alk1, alk2, alk3, alk4}, we do not try to
fix the gauge invariance of the minimally coupled action (when the
metric is perturbed away from the background AdS), instead we consider
a different problem, that of tree-level unitarity.

This issue was first discussed in \cite{Cucchieri:1994tx} in the
context of massive fields in 4d Minkowski space. There, tree-level Compton scattering amplitudes of gravitons
with massive HS fields were considered.

There are two kinds of
Compton scattering amplitudes. One is known as a ``sea-gull'' diagram,
which is just a four-point vertex. The other contains an HS
propagator. Only the second type potentially leads to violation of
tree-level unitarity. It can schematically be written as $J\Pi J$,
where
\be\nonumber
J\equiv\frac{\delta S(\phi,e^a_\mu)}{\delta \phi}
\ee
is the current associated with the graviton scattering. Here, $S$ is the action of a generic field $\phi$ and
$e^a_\mu$ is the vier-bein of the perturbed spacetime. Finally, $\Pi$
is the propagator of the HS field on the unperturbed spacetime. $\Pi$
contains $\frac{1}{m^2}$ terms. Generically, if $J$ is independent of
$m$, these terms in the propagator give rise to
$O\left(\frac{s^2}{m^2M_{Pl}^2}\right)$ terms in the scattering amplitude, where the
$\frac{1}{M_{Pl}^2}$ comes from the two gravitons in the currents. $O\left(\frac{s^2}{m^2M_{Pl}^2}\right)$
terms are dangerous as they violate the unitarity bound at
$\sqrt{s}\sim\sqrt{mM_{Pl}}$, {\it i.e.} well below the Planck scale. An obvious way to
solve this problem is to have a mass-dependent current of the form
$J\sim mX$, where $X$ has a finite $m\rightarrow 0$ limit.

More specifically, it is the components of $\Pi$ with all indices in
pure gauge directions (longitudinal indices) that contain the
dangerous $\frac{1}{m^2}$ terms: if $\mathring{\phi}$ is a pure gauge
field ($\mathring{\phi}\equiv \partial \epsilon$) then $\Pi\cdot\mathring{\phi}=m^{-2}\mathring{\phi}$
\footnote{Here the gauge transformation we are
referring to is that under which the free massless HS action is invariant. We suppress indices for simplicity.}. Therefore we only need
\be\nonumber
\int \partial \epsilon\cdot \tilde J =mY\neq 0\ ,
\ee
where again $Y$ is finite in the $m\rightarrow 0$ limit and $\tilde J$ is the current density.
Here $\int \partial \epsilon\cdot \tilde J\sim \int \epsilon \cdot (\partial \cdot \tilde J)$
cannot vanish due to the soft breaking of gauge invariance by the mass $m$.

We note that
\be\nonumber
\int \partial \epsilon\cdot \tilde J=\delta \phi\cdot \frac{\delta S(\phi,e^a_\mu)}{\delta \phi}=\delta S(\phi,e^a_\mu)\ ,
\ee
where $\delta S(\phi,e^a_\mu)$ is the on-shell gauge variation of $S$.
In other words, tree-level unitarity is obtained by requiring that the
gauge variation of $S$ with all fields on-shell (using the free
equations of motion) and linearised in
the graviton is of the form $mY$. In \cite{Cucchieri:1994tx}
non-minimal terms proportional to inverse powers of $m$
are added to the action to achieve this. In this way, although tree-level unitarity
is restored up to $O(h^2)$ terms,
gauge invariance is still softly broken
on-shell due to the mass term. The analysis of \cite{Cucchieri:1994tx}
deals with the unitarity of the longitudinal
contributions to the tree-level scattering amplitudes, but does not
deal with that of the transverse contributions which are still present
even in the massless limit. By contrast the massless theory does not
have these transverse contributions and this discrepancy is due to a
similar discontinuity as that of the massive graviton propagator in
the massless limit \cite{vdv}.

In the case of massless fields in $AdS_5$, the cosmological constant
acts as a mass and so tree-level unitarity is similarly violated in the Compton scattering amplitude of gravitons
with HS fields at
$\sqrt{s}\sim\sqrt{\sqrt{\Lambda}M_{Pl}}$. This problem can be avoided again by adding
non-minimal terms to the action, as was done in \cite{GK} in the fermionic case. Moreover, one has
the extra bonus of having tree-level conserved currents, so that the
dangerous $J\Pi J\Big|_{\tiny longitudinal}$ vanishes on shell. This means that the
non-minimal terms in fact restore the ``on-shell gauge-invariance'', in other words, the
gauge variation of the interaction action is zero once the free equations of motion and gauge constraints have been imposed.
Also, thanks to the fact that higher spins are massless in $AdS$,
one has that $J\Pi J\Big|_{\tiny transverse}=0$ (the propagator does not contain transverse directions, in contrast
to the massive case).
Tree-level unitarity of {\it all of} the Compton scattering amplitude is therefore restored, up to $O(h^2)$ terms,
thanks to the addition of non-minimal terms to the action. In fact, in this case,
the only non-vanishing contributions to the tree-level Compton scattering amplitude are
the tree-level unitary sea-gull diagrams.

In the first part of this paper we
correct the non-minimal terms introduced in \cite{GK} to restore the tree-level unitarity of the
Compton scattering of gravitons with fermionic HS fields and extend
the analysis to the bosonic case.

In the second part of the paper, we consider massless HS fields living
in a braneworld scenario. We show that Standard Model (SM) particles,
living on a three-dimensional brane,
can co-exist with HS fields without spoiling the HS on-shell gauge
invariance.
This co-existence is due to the fact that HS interact
only gravitationally with SM fields on a brane. HS fields might then be possible Dark Matter (DM) candidates.
However, as we shall show, HS particle
production by graviton scattering (the only bulk
interaction considered here) is only important at trans-planckian
scales, and so if the hypothesis of HS being DM candidates is to be tested
further, a new
mechanism to explain their current observed abundances must be found. We leave this for future research.

\subsection{Higher-spin fields}

In the following, in order to be as general as possible, we will work on an effective theory of HS
(for string theory modes decaying into standard
model particles see \cite{iengo}).

A generic massless bosonic particle of integer spin $s$ is described by a  totally symmetric
tensor of rank $s$, $\Phi_{\mu_1\mu_2...\mu_s}$, while a fermionic
particle of spin $s$  is described by a totally
symmetric tensor-spinor of rank $s-\frac{1}{2}$,
$\Psi_{\mu_1\mu_2...\mu_{s-\frac{1}{2}}}$~\footnote{Here we consider
  only totally symmetric HS fields. It is possible that the reduction
  of string theory to our scenario may also give rise to mixed
  symmetry HS fields~\cite{tsusag, resh3, resh4, ouv1}. We leave the
  consideration of the phenomenological implications of these fields
  to future work.}.
These fields are defined up to gauge transformations and they are subject to certain
constraints such that the corresponding theories are ghost free.
This means that they describe exactly two propagating
modes of $\pm s$.

It is known that there is no problem of writing down HS field
equations in flat space for free fields. The problems appear when one
considers interactions of these fields. The most obvious
interaction is the gravitational interaction.  An immediate way
of introducing the latter is to replace ordinary derivatives with
covariant ones in order to maintain general covariance. However, with
this replacement gauge invariance is lost: to prove gauge invariance
in flat space one needs to commute derivatives and their lack
commutativity now in curved space leads to hard gauge breaking terms proportional to the Weyl tensor of the spacetime~\cite{Aragone:1979hx}.
This means
that HS fields minimally coupled to gravity have acausal propagation in curved
spacetimes and cannot consistently be defined. This ``no-go theorem'' can however be circumvented
on spacetimes with
vanishing Weyl tensor, i.e. on conformally flat space-times, such
as de Sitter ($dS$) and Anti-de Sitter ($AdS$) spacetimes~\cite{Deser}. Indeed,
soon after the results of
\cite{Fronsdal:1978rb},\cite{Fang:1978wz}, propagation of
HS fields on (A)dS were discussed
in~\cite{Fronsdal:1978vb}.
In particular, by  gauging  an infinite-dimensional generalisation of the target
space Lorentz algebra, consistent interactions
of HS fields were introduced~\cite{Vasss},\cite{Sezgin}. Such consistent interactions do not have
a flat space limit as they are based on a generally covariant curvature expansion on $(A)dS$ spacetime with
expansion parameter proportional to the $(A)dS$ length.

In this paper, we will discuss both HS fields living in unbounded
$AdS$ spacetime and in bounded $AdS$ spacetime. The particular bounded
$AdS$ spacetime we will consider is the Randall-Sundrum scenario~\cite{Randall:1999ee},
which has been extensively studied as an alternative to
compactification and in connection with the hierarchy
problem~\cite{AK}. In this scenario SM fields are assumed to live on
the boundary of the $AdS$ space, a braneworld.

\subsection{Higher spins in a Randall-Sundrum scenario}\label{hsinRS}

In this section we set up the notation.
Recall that $AdS$ is a maximally symmetric spacetime, and in
Gaussian-normal coordinates, its metric takes the form
\be
ds^2=e^{-2\sigma}\eta_{ab}dx^a dx^b+dy^2\ , \label{metric}
\ee
where $a,b,...=0,...,3$, $y=x_5$ \footnote{Throughout
  the paper we use greek letters to denote indices taking values
  $0,\ldots,4$, and latin indices for indices taking values $0,\ldots,3$.} and $\sigma=\sqrt{-\Lambda}/2 \, y$ where $\Lambda$ is the spacetime cosmological constant, $\sigma$ is called
the warp factor.
A Randall-Sundrum II (RSII) spacetime~\cite{Randall:1999ee} is a $Z_2$
orbifold of $AdS$, and thus, its metric is (\ref{metric}) with $\sigma=2 a |y|$ and $a=\sqrt{-\Lambda}/4$.
In RSI, there exists an ``end of the
world"  at $y=\pi R$ so that $0\leq y\leq \pi R$.

The second derivative of $\sigma$ appears in the curvature tensors
producing
$\delta$-function contributions to both Riemann and Ricci tensors. These contributions may be canceled
by putting branes of appropriate fine-tuned tensions at the fixed points of the $Z_2$ orbifold. The branes are 4D flat
Minkowski spacetimes and they are the boundaries of the
bulk $AdS$ background.
The boundary at $y=0$ is the
UV brane while the brane at $y=\pi R$ is the IR one.  In~\cite{Randall:1999ee}, our
Universe is on the UV brane. This model is considered a valid
alternative to compactification and therefore we will use it as our framework.

In curved spacetime, one has to modify the definition of the spacetime covariant derivative
in order to maintain a local Lorentz invariance
of a semi-integer spin field \cite{GK}. In $AdS_5$, and in Gaussian-normal coordinates, this is achieved by introducing the covariant
derivative $\bar
\nabla_\mu=\nabla_\mu+\frac{1}{2}\gamma_5\gamma_a\ \sigma'\delta^a_\mu+a\gamma_\mu$.

A central issue when boundaries are present, as in RS,  is the
boundary condition problem. In varying the action to find the HS
fields' equations of motion, boundary terms are generated which must
vanish independently from the bulk terms, and so appropriate boundary
conditions must be introduced. For fermionic fields,
the action is of the form
\be
S= \frac{1}{2}\int d^5 x\sqrt{-g}\bar\Psi_{\alpha_1...\alpha_{s-1/2}}\gamma^\beta\bar\nabla_{\beta}\Psi^{\alpha_1...\alpha_{s-1/2}}+...\ ,
\ee
where the $\ldots$ indicate more terms that do not affect the boundary conditions.
In the presence of boundaries, the boundary terms generated by the
variation of the above action can be made to
vanish by imposing
\be
\left(\delta \Psi^L\cdot\Psi^R-\delta \Psi^R\cdot
\Psi^L\right)\Big|_{0,\pi R}=0\ ,
\label{fermbdycond}
\ee
where $L,R$ stands for the chiral left and right projections.

As we are interested in the  $Z_2$ symmetry $y\to -y$, it is
easy to see that the action $S$ is $Z_2$ symmetric if $\Psi(-y)=\pm \gamma^5 \Psi(y)$.
We can take the positive sign without loss of generality. This means that the right-handed
field will in general have a ``kink'' profile
around $y=0$, and the boundary condition (\ref{fermbdycond}) reduces to
\be
\Psi_L^+=0,
\nonumber
\ee
\be
\Psi_R^+=0,\nonumber
\ee
or
\be
\Psi_L^+=\Psi_R^+.
\label{fermbdycondexplicit}
\ee

With a similar procedure, we can consider a bosonic field with action
\begin{eqnarray}
S=\frac{1}{2} \int d^5 x\sqrt{-g}\nabla_{ \mu}\Phi_{ \alpha_1...\alpha_{s}}\nabla^{\mu}\Phi^{ \alpha_1...\alpha_{s} }
+&2a(s-1)\int d^5x\sqrt{-g}\delta(y)\Phi_{\alpha_1..\alpha_s}\Phi^{\alpha_1..\alpha_s}+...\ ,\label{bos0}
\end{eqnarray}
where the second term must be introduced to restore gauge invariance
in a spacetime with boundaries \cite{GK} \footnote{Note that the factor in front of the boundary action is $(s-1)$ and not $(s-2)$. This
correct a typo in \cite{GK}}, and the $\ldots$ again
indicate terms which do not affect the boundary conditions.

Without the $Z_2$-symmetry, the variational principle, in gaussian-normal coordinates,
is well defined if
$\left(\delta \Phi\cdot n^a\d_a\Phi\right)\Big|_{0,\pi R}=0$.
However as the spacetime is $Z_2$ symmetric, the bulk field variation has a term like
$\delta \Phi_{\alpha_1...\alpha_{s}}\nabla_\mu\nabla^{\mu}\Phi^{ \alpha_1...\alpha_{s}}$,
which in fact contains a boundary term on the fixed points of the spacetime.
This happens because the second derivative of the metric contains
delta-functions peaking at the fixed points.
Then, in gaussian-normal coordinates in an $AdS$ spacetime,
we obtain the following two possible boundary conditions for a bosonic field $\Phi$ of any spin \cite{GK}
\paragraph{\rm Robin:} $\Phi'(y)-4a(s-1)\Phi(y)\Big|_{0,\pi R}=0\ ,$
\paragraph{\rm Dirichlet:} $\Phi(y)\Big|_{0,\pi R}=0$ .

As we will discuss later, only the Dirichlet boundary conditions are allowed by gauge invariance if
matter is present on the brane.

Finally, we need to discuss gauge constraints. Gauge-invariant HS fields
are realised whenever the HS action is invariant under the following gauge transformations \cite{GK},

{\rm Fermionic:}
\be
\delta\Psi_{\alpha_1...\alpha_{s-1/2}}=\bar\nabla_{(\alpha_1}\epsilon_{\alpha_2...\alpha_{s-1/2})}\ ,
\label{generalvar}
\ee

{\rm Bosonic:}
\be \label{df}
&&\delta \Phi_{\mu_1\mu_2...\mu_s}=\nabla_{(\mu_1}\xi_{\mu_2\mu_3...\mu_s)}\ .
\ee
The bosonic fields are double traceless symmetric tensors, while the
fermionic fields are triple $\gamma$-traceless symmetric
tensor-spinors. In five dimensions a massless spin s particle has
$2s+1$ degrees of freedom. Therefore, we need to impose gauge
constraints on $\Phi$ and $\Psi$ to eliminate unphysical degrees of
freedom. It turns out that on-shell we are allowed to impose
\be
\Phi^{\mu}\!_{\mu\mu_3\ldots\mu_s}=0
\label{bgauge}
\ee
and
\be
\gamma^{\mu}\Psi_{\mu\mu_2\ldots\mu_{s-\frac{1}{2}}}=0.
\label{fgauge}
\ee
So that the gauge transformations preserve these constraints we also
impose $\gamma^{\mu}\epsilon_{\mu\mu_3\ldots\mu_{s-\frac{1}{2}}}=0$,
$\nabla^{\mu}\epsilon_{\mu\mu_3\ldots\mu_{s-\frac{1}{2}}}=0$,
$\xi^{\mu}\!_{\mu\mu_3\ldots\mu_s}=0$ and $\nabla^{\mu}\xi_{\mu\mu_3\ldots\mu_s}=0$.

\section{Tree-level unitary actions in AdS}\label{AdS}

We have already noted in the introduction that in general the gauge
invariance in Minkowski space is lost when the spacetime becomes
curved and the HS are
minimally coupled to gravity. Instead of a general spacetime, let us
consider a perturbation away from flat space and hence the
interactions between gravitons (the metric perturbations away from
flat space) and HS. The gauge breaking terms are proportional to the
Riemann tensor.
These terms are non-zero even for on-shell gravitons and therefore
tree-level unitarity for the
graviton-HS scattering amplitudes is
unavoidably lost \cite{Cucchieri:1994tx}.
The situation is different for massive HS fields. In this case,
without introducing additional gauging, the gauge breaking terms of
the massless theory can be cancelled by
a non-minimal interaction like $\frac{1}{m}\Phi_{\alpha\beta...} {\cal R}^{\alpha\mu\nu\beta} \Phi_{\mu\nu...}$~\cite{Porrati}.
This interaction cancels hard gauge-breaking terms, {\it i.e.} terms that do not vanish
in the massless limit, although gauge invariance is still softly broken due to an explicit mass term.
Hence, tree-level unitarity
of
(the longitudinal parts of)
the Compton scattering
amplitudes
is restored up to the Planck scale~\cite{Cucchieri:1994tx},\cite{Porrati}.
The price paid is the violation of the equivalence principle due to the introduction of the
non-minimal interaction terms~\cite{Porrati},\cite{Giannakis}. Although such terms look odd,
experience from electromagnetic interactions
suggests that the physical requirement is tree-level unitarity~\cite{Tiktopoulos},\cite{Ferrara}
rather than minimal coupling. It is clear of course that the massless
limit for this theory is not defined.

In $AdS$, HS actions naturally contain a non-derivative term proportional to the cosmological
constant. This is something like
having a mass term in the Minkowskian
case. If now, as in the flat space case, we perturb away from pure $AdS$
(where we know the HS minimal action is gauge-invariant), then it has
been shown \cite{GK}, that for fermionic fields, a non-minimal interaction with gravity can cancel gauge breaking terms proportional to the Riemann
tensor. In this section, we correct the non-minimal interaction
proposed by \cite{GK} and show that a similar non-minimal interaction
can be found for bosonic fields. Unlike the mass term in the
four-dimensional action in a Minkowski background, in our case of
$AdS$ the
non-derivative term does not break gauge invariance. Therefore just the
cancellation of the hard gauge-breaking terms restores the
on-shell
gauge invariance of
the interacting theory.
Non-linear gauge invariance might then be restored as an
infinite series of this kind of non-minimal interactions \cite{Vass,Vasss}.

Technically, the
on-shell
gauge invariance is obtained in our method if and only if the gravitational background in which the
HS is propagating is a constant curvature background and the higher
spin field is on-shell. As is sketched in the appendix, by a lengthy computation it can be
shown that
the following non-minimal actions couple HS fields consistently to linear gravity
under the gauge transformations (\ref{generalvar},\ref{df}) at tree-level on an $AdS_5$ background:
\paragraph{Fermionic}
\be
S^f=S_0^f+S^f_{nm}+\Delta S^f\label{fermionic}
\ee
where
\begin{eqnarray}
S_0^f=\int d^5x
\sqrt{-g}\Big{[}\!-\!\frac{1}{2}\bar{\Psi}_{\alpha_1...\alpha_{s-1/2}}Q^{\alpha_1...\alpha_{s-1/2}}
&+&\frac{1}{4}(s\!-\!\frac{1}{2})\bar{\Psi}_{\mu\rho\alpha_3...\alpha_{s-1/2}}
\gamma^\rho\gamma_\sigma Q^{\sigma\mu\alpha_3...\alpha_{s-1/2}}
\cr &+&\frac{1}{8}(s\!-\!\frac{1}{2})(s\!-\!\frac{3}{2})\bar{\Psi}^\mu{}_{\mu\alpha_3...\alpha_{s-1/2}}
Q^\nu{}_\nu{}^{\alpha_3...\alpha_{s-1/2}} \Big{]}\ , \label{as}
\end{eqnarray}
\cite{Fang:1978wz, deWitFreedmanetc, Metsaev:2004ee, Metsaev:2005ws,
  Metsaev:2007rw, resh2, ouv2, tom1}, and the non-minimal interactions
\be\label{sf}
S^f_{nm}&=&\frac{3(s-\frac{3}{2})(s-\frac{1}{2})}{16as}\int
d^5\!x \sqrt{-g}\bar{\Psi}_{\mu\nu\alpha_3\ldots\alpha_{s-\frac{1}{2}}}
\mathcal{W}^{\mu\rho\nu\sigma}\Psi_{\rho\sigma}\!^{\alpha_3\ldots\alpha_{s-\frac{1}{2}}}
\ee
\be \label{DSF}
\Delta S^f=-\frac{3(s-\frac{3}{2})(s-\frac{1}{2})(s-\frac{5}{2})}{16as}\int
d^5\!x \sqrt{-g}\bar{\Psi}^{\lambda}\!_{\mu\nu\alpha_4\ldots\alpha_{s-\frac{1}{2}}}\gamma_{\lambda}\left(\bar{\nabla}_{\alpha_3}
\mathcal{W}^{\mu\rho\nu\sigma}\right)\Psi_{\rho\sigma}\!^{\alpha_3\ldots\alpha_{s-\frac{1}{2}}}\ .
\ee
We have used the notation
\be
Q_{\alpha_1...\alpha_{s-1/2}}=\gamma^\rho\bar\nabla_{\ \rho}\Psi_{\ \alpha_1\ ...\alpha_{s-1/2}}-
 \gamma^\rho\bar\nabla_{(\alpha_1}\Psi_{\ \alpha_2\ ...\alpha_{s-1/2})\rho}
 +2a (2s-3)
\Psi_{\alpha_1...\alpha_{s-1/2}}\ ,
\ee
and
\be
\mathcal{W}^{\mu\rho\nu\sigma}=W^{\mu\rho\nu\sigma}-\frac{1}{2}W_{\alpha\beta}\!^{\rho\mu}\gamma^{\nu\sigma\alpha\beta}+\frac{1}{6}
W_{\alpha\beta}\!^{\rho\mu}\gamma^{\alpha\beta}g^{\nu\sigma}-
\frac{1}{3}W_{\alpha\beta}\!^{\rho\mu}\gamma^{\nu\alpha}g^{\beta\sigma}-\frac{1}{3}W_{\alpha}\!^{\nu\rho\mu}\gamma^{\sigma\alpha}\ ,\label{4.6}
\ee
where $W_{\mu\nu\alpha\beta}$ is the spacetime Weyl tensor. Note that (\ref{4.6}) corrects the corresponding formula in \cite{GK}.

The second part of (\ref{DSF}) arises only for spin $s>5/2$. It
vanishes on-shell but gives a non zero contribution to the gauge variation
{\it s.t.} the total action's gauge variation can be eliminated by a
field redefinition as in \cite{Cucchieri:1994tx}.
\paragraph{Bosonic}
In this case we have
\be
S^b=S^b_0+S^b_{nm}+\Delta S^b\ ,
\ee
where
\begin{eqnarray}\label{zero}
S^b_0\!\!&=&-\!\!\int d^5x \sqrt{-g}\left(
\frac{1}{2}\nabla_\mu\Phi_{\alpha_1...\alpha_s}\nabla^\mu\Phi^{\alpha_1...\alpha_s}-\frac{1}{2}s
\nabla_\mu{\Phi^\mu}_{\alpha_2...\alpha_s}\nabla_\nu\Phi^{\nu\alpha_2...\alpha_s}\right.+\frac{1}{2}s(s\!-\!1)
\nabla_\mu{\Phi^\nu}_{\nu\alpha_3...\alpha_s}\nabla^\kappa{\Phi_\kappa}^{\mu\alpha_3...\alpha_s}
\cr &-&\frac{1}{4}s(s\!-\!1)\nabla_\mu{\Phi^\kappa}_{\kappa\alpha_2...\alpha_s}\nabla^\mu
{\Phi_\lambda}^{\lambda\alpha_2...\alpha_s}-\frac{1}{8}s(s\!-\!1)(s\!-\!2)
\nabla_\mu{\Phi^{\mu\kappa}}_{\kappa\alpha_4...\alpha_s}\nabla^\nu
{\Phi_{\nu\lambda}}^{\lambda\alpha_4...\alpha_s}+\\
&+&2a^2\left(s^2\!-s\!-\!4)\right)
\Phi_{\alpha_1...\alpha_s}\Phi^{\alpha_1...\alpha_s} \!-\! a^2 s(s\!-\!1)\!
\left(s^2\!+\! s\!-4\right){\Phi^\mu}_{\mu\alpha_2...\alpha_s}{\Phi_\nu}^{\nu\alpha_2...\alpha_s} \!\!\!
\left. \phantom{\frac{1}{2}}\!\!\!\! \right)\ , \nonumber
\end{eqnarray}
\cite{Fronsdal:1978vb, resh1, buchbinderpashnevtsulaia, tom1}, and the non-minimal interactions are
\be\label{snm}
S_{nm}^b=\frac{1}{2}s(s-1)\int d^5x\sqrt{-g}\Phi_{\alpha\gamma}{}^{\mu_3...\mu_s}W^{\alpha\beta\gamma\delta}\Phi_{\beta\delta\mu_3...\mu_s}\ ,
\ee
and
\begin{eqnarray}\label{DSB}
\Delta S^b&=&\frac{s(s-1)(s-2)}{8a^2(s^2-s-4)}\int d^5x \sqrt{-g}\left[\nabla^\lambda\Phi_{\alpha\gamma\lambda}{}^{\mu_4...\mu_s}
\nabla^\rho W^{\alpha\beta\gamma\delta}\Phi_{\beta\delta\rho\mu_4...\mu_s}-\Phi_{\alpha\gamma\lambda}{}^{\mu_4...\mu_s}W^{\alpha\beta\gamma\delta}
\nabla^\lambda\nabla^\rho\Phi_{\beta\delta\rho\mu_4...\mu_s}\right]\cr
&+&\frac{s(s-1)(s-2)}{4a^2(s^2-s-4)}\int d^5x \sqrt{-g}\left[\Phi^\nu{}_{\nu\alpha}{}^{\mu_4...\mu_s}\left(\nabla^\rho W^{\alpha\beta\gamma\delta}
\right)\nabla_\gamma\Phi_{\rho\beta\delta\mu_4...\mu_s}\right]\cr
&+&\frac{s(s-1)(s-2)(s-3)}{8a^2(s^2-s-4)}\int d^5x \sqrt{-g}\left[\Phi^\nu{}_{\nu\alpha\gamma}{}^{\mu_5...\mu_s}\left(\nabla^\lambda\nabla^\rho
W^{\alpha\beta\gamma\delta}\right)\Phi_{\beta\delta\lambda\rho\mu_5...\mu_s}\right]\ .
\end{eqnarray}
The third part of (\ref{DSB}) arises only for spin $s>3$.
The action (\ref{DSB}) vanishes on-shell but gives non-zero
contributions to the gauge variation {\it s.t.} again the gauge variation of
the total action can be eliminated by a field redefinition as in \cite{Cucchieri:1994tx}.

We note also that the non-minimal part, $S_{nm}$, is
analytic
in the limit $a\rightarrow 0$. This is correct as one can show that the action
$S_0+S_{nm}$
is actually equivalent to a minimal action \cite{AragoneLa},
plus terms proportional to the Ricci tensor that vanish when the
graviton is put on-shell \cite{Cucchieri:1994tx}.

At first sight this action may look odd to a reader familiar with the results of Fradkin and Vasiliev
\cite{fradvas} in 4d, as here we have a maximum of four derivatives while \cite{fradvas} have 2s-2. This
difference is due to us only being interested in tree level unitarity. To restore the full ``off-shell''
gauge invariance at the linearised level, higher derivatives should appear \footnote{We thank Nicolas
Boulanger and Per Anders Sundell for pointing this out to us.}. The
two theories are therefore inequivalent and both have advantages and disadvantages. The theory presented
here has only two derivatives on-shell so that the usual field theory observables (energy, momentum, etc.) can be straightforwardly used. Nevertheless,
although the only physical requirement of unitarity has been fulfilled, ``off-shell'' gauge invariance
is broken by the gravitons in our theory. The theory of \cite{fradvas} has instead the nice mathematical
feature of being ``off-shell'' gauge invariant, and unitary, at linearised level.
Nonetheless, the appearance of higher-derivatives in \cite{fradvas} in the equations of motion of
the higher spin fields is problematic for a canonical approach to physical observables.

\section{HS and brane SM}\label{HS-SM}

Let us now consider placing a brane in the AdS spacetime with
$Z_2$-symmetry across it. The metric is then (\ref{metric})
with $\sigma=2a|y|$. The extrinsic curvature is not continuous across
$y=0$ (the brane location) and this gives rise to terms proportional
to $\delta(y)$ in the five dimensional curvature
tensors \cite{gr-qc/0004021}.

The actions derived in section \ref{AdS} are in general no longer
gauge-invariant in this spacetime. Indeed, a fundamental ingredient in deriving their gauge
invariance was the fact that gravity was propagating in a constant
curvature spacetime. In the case of a vacuum brane with
\be
R_{\alpha\beta}=\Lambda g_{\alpha\beta}-\frac{1}{3}\lambda q_{ab}\delta^a_\alpha\delta^b_\beta\delta(y),
\ee
where $\lambda$ is the brane tension and $q_{ab}$ is the induced metric on the brane, however, gauge invariance can be
restored by simply adding boundary mass terms for HS \cite{GK}. This
is because, in the RS background the Ricci tensor is still
proportional to the metric. In contrast, in the general case with
arbitrary matter on the brane, we instead have
\be
R_{\alpha\beta}=\Lambda g_{\alpha\beta}+\left[T_{ab}-\frac{1}{3}\left(T+\lambda\right)q_{ab}\right]\delta(y)\delta^a_\alpha\delta^b_\beta\ ,
\ee
where $T_{ab}$ is the energy momentum tensor of the brane's matter fields.

In this case the Ricci tensor is not proportional to the metric and
therefore, although gauge invariance is
not in general spoiled in the bulk, it might be on the brane.

Finding suitable non-minimal actions for the fields in this case can
be approached in one of two ways. One way is to repeat the analysis of
section \ref{AdS}, just with a different metric ((\ref{metric})
with $\sigma=2a|y|$). This is the approach we adopt. Another approach
would be to regard RS as two AdS spaces with boundaries glued together
at the brane, and look for the Gibbons-Hawking \cite{GH}-like boundary terms of the AdS
non-minimal actions that we have already found. We adopt the first
approach because we find it conceptually simpler, but the second
approach would be equally valid.

Let us firstly consider
bosonic higher spin fields on an RS background.
For simplicity we will restrict our attention to fields which obey
the gauge restriction $\Phi_{5...}=0$ \footnote{This constraint removes lower spin fields
  from the dimensionally reduced theory on the brane.}, but our
conclusions are general. The variation
of the minimal action (\ref{zero}) is
\begin{eqnarray}\label{nonads}
\delta S_0^b=\int d^5x\sqrt{-g} \xi_{a_2\ldots a_s}\{\nabla^{\mu}\left(R_{b \mu}\Phi^{b a_2\ldots a_s}\right)
+2(s-1)\nabla^{\mu}\left(R^{a_2}{}_{b a_1 \mu}\Phi^{a_1 b\ldots a_3}\right)-(s-1)\nabla^{\mu}R^{a_2}{}_{b a_1 \mu}\Phi^{a_1\beta\ldots a_3}\}\ .
\end{eqnarray}
These terms can be split into bulk and boundary parts. The bulk
parts have already appeared in the no boundary case and so can be
cancelled with the same non minimal terms introduced
before. Therefore, we only need to consider the boundary parts. We can
write these in terms of the brane's extrinsic curvature,
$K_{ab}\Big|_{y=0}=\tilde K_{ab}\propto
T_{ab}-\frac{1}{3}(T+\lambda)q_{ab}$\cite{gr-qc/0004021}. Using the
Codacci equations, we obtain the
following boundary action:
\be\label{boundary}
\delta S_{0,\mbox{\tiny bound.}}^b=\int d^4x\sqrt{-q} \xi_{a_2\ldots
  a_s}\{\nabla^m \left(\tilde K_{m b}\Phi^{b a_2\ldots a_s}\right)-(s-1)
\nabla^{a_2}\tilde K_{b a_1}\Phi^{a_1ba_3\ldots a_s}+(s-1)\nabla_b \tilde K_{a_1}{}^{a_2}\Phi^{a_1 ba_3\ldots a_s}\}\ .
\ee
The only terms that could be added to the action to cancel this are of
the form
\[
S_{counter}\propto \int dx^4\sqrt{-q} \Phi\cdot \tilde K_{mn}\cdot\Phi\ ,
\]
with a suitable contraction of indices, or
\[
S_{counter}\propto \int dx^4\sqrt{-q} \Phi\cdot \tilde K\cdot\Phi\ .
\]
One can check that not all of the terms of (\ref{boundary}) can be
simultaneously cancelled by the gauge variation of such terms.
This no-go result is circumvented only in a pure tension brane in which $\tilde K_{ab}\propto q_{ab}$.
In this case the boundary action to add is
\be
S_{counter}=2a(s-1)\int dx^4\sqrt{-q} \Phi_{a_1\ldots a_s}\Phi^{a_1\ldots a_s}\ ,
\ee
as given in section \ref{hsinRS}. Obviously this no go result does not apply to the graviton in which the boundary terms are automatically removed by the linearized
Einstein equations.

To conclude then, it is not possible to construct a linearised
  on-shell gauge-invariant
  action for the bosonic higher-spin fields in perturbed RS unless we
  impose $\Phi_{\mu_1\ldots\mu_s}=0$ on the brane. In this
  case a brane observer could not observe bosonic higher spins. Note however that gauge invariance is
preserved, without any need for additional boundary conditions, for any spin $s\leq 2$.

Let us now turn our attention to fermionic higher spin fields. The
fermionic case differs significantly from the bosonic in that it
{\it is} possible to construct an
on-shell
gauge-invariant action in perturbed RS
without imposing any additional boundary conditions on the higher spin
fields. This means that a brane observer {\it may} measure fermionic HS by their bulk projection onto
the brane.

To be more precise one can show that for a general background the
on-shell
gauge-invariant action $S_T=S^f_0+\sum_{i=1}^3\Delta S_i^f$ is
\begin{eqnarray}
\Delta S_1&=&
\frac{3(s-\frac{1}{2})(s-\frac{3}{2})}{8a(s-\frac{3}{4})}\int
d^5x\sqrt{-g}\bar{\Psi}_{\mu\nu\alpha_3\ldots\alpha_s}{\cal W}^{\mu\rho\nu\sigma}\Psi_{\rho\sigma}\!^{\alpha_3\ldots\alpha_s},\nonumber\\\Delta
  S_2&=&-\frac{2s(s-\frac{1}{2})}{3a(1-4s)}\int
d^5x\sqrt{-g}\left[\bar{\Psi}^{\rho}\!_{\alpha_2\ldots\alpha_s}R_{\rho\tau}\Psi^{\tau\alpha_2\ldots\alpha_s}+
\bar{\Psi}^{\rho}\!_{\alpha_2\ldots\alpha_s}R_{\tau\alpha}\gamma^{\alpha}\!_{\rho}\Psi^{\tau\alpha_2\ldots\alpha_s}\right]\nonumber\\\Delta
S_3&=&\frac{(s-\frac{1}{2})}{2a}\int
d^5x\sqrt{-g}\bar{\Psi}^{\lambda}_{\mu\nu\alpha_4\ldots\alpha_s}\gamma_{\lambda}\left[\frac{(s-\frac{3}{2})}{a(s-\frac{3}{4})}\nabla_{\sigma}
W_{\alpha}\!^{\sigma\rho\mu}\Psi_{\rho}\!^{\alpha\nu\alpha_4\ldots\alpha_s}-
\frac{3(s-\frac{3}{2})(s-\frac{5}{2})}{8a(s-\frac{3}{4})}\bar{\nabla}_{\alpha_3}{\cal W}^{\mu\rho\nu\sigma}
\Psi_{\rho\sigma}\!^{\alpha_3\ldots\alpha_s}\right.\nonumber\\&&+
\frac{2s}{3a(1-4s)}(D^{\rho}R_{\rho\tau})\Psi^{\tau\mu\nu\alpha_4\ldots\alpha_s}+
\frac{2s}{3a(1-4s)}\gamma^{\alpha}\!_{\rho}(D^{\rho}R_{\tau\alpha})\Psi^{\tau\mu\nu\alpha_4\ldots\alpha_s}\nonumber\\&&+
\left.\frac{4s(s-\frac{3}{2})}{3a(1-4s)}(D^{\alpha_2}R^{\mu}_{\tau})\Psi^{\tau\nu\alpha_2\alpha_4\ldots\alpha_s}\right].
\end{eqnarray}
On-shell, the boundary contribution of each term in $\sum_{i=1}^3\Delta S_i^f$ is of the form
\be
\int d^4 x\sqrt{-q}
\bar{\Psi}\cdot\gamma\cdots\gamma\cdot K\cdot\Psi,
\ee
with some contraction of indices, and an even number of $\gamma$
matrices, where $K_{\mu\nu}$ is the extrinsic curvature of the
brane. The contribution to the brane action is then
\be
\int d^4 x\sqrt{-q}\left[
\bar{\Psi}\cdot\gamma\cdots\gamma\cdot K\cdot\Psi\right]^+_-,
\label{fermionicbdy}
\ee
where $+(-)$ indicates the value of the expression on the $y>0(y<0)$
side of the brane. It can be shown that for all the terms in $\sum_{i=1}^3\Delta S_i^f$ (and its gauge variation) the
indices of the $\gamma$ matrices in (\ref{fermionicbdy}) are
restricted to the brane directions, and so for all of the possible
boundary conditions in (\ref{fermbdycondexplicit}), (\ref{fermionicbdy}) vanishes (we remind the reader that $K^+_{\mu\nu}=-K^-_{\mu\nu}$
across the brane).
This implies that no interaction of HS and SM is possible, not even
via higher dimensional gravity mediation \footnote{This corrects the
  erroneous claim of \cite{GK} that such interaction is possible.}. This can be physically understood by considering the Weinberg-Witten result
\cite{WW} which forbids any gauge invariant interactions with HS fields and matter fields in a flat background. This indeed implies that the effective
HS theory on the brane cannot possibly have any interaction with matter fields localized on the brane. Therefore the only possibility is that the HS
fields do not interact with the brane. This automatically happens for fermionic fields thanks to their kink profile across the brane. For bosonic
field the non-interaction must be instead imposed by the Dirichlet boundary condition.

Interaction of HS fields on the brane with brane gravitons is, however, still possible, as
$S_0^f$ {\it does} contribute to the brane action. We discuss this
possibility in the next section.

\section{HS production by graviton scattering}

In the RSII scenario, when we dimensionally reduce the HS field to the
brane, we obtain a massless mode and a continuum of massive Kaluza-Klein
(KK) modes, but with no mass gap \cite{GK}. This continuum of massive
modes gives rise to `unparticles' \cite{giorgi}. We shall see below that the cross-sections for gravitons to decay into
HS unparticles is only non-negligible at trans-planckian scales. This
means that a brane observer will experience stable higher spin
unparticles for the whole evolution of the classical
Universe ({\it i.e.} far from the quantum era), and so HS unparticles
might be used as Dark matter candidates
\footnote{The idea of using unparticles (although not HS unparticles) as DM candidates was first
  proposed by \cite{kik}.}.

To prove the stability of HS up to Planck scale, we will consider for simplicity only the massless KK modes for both the
fermionic HS and the higher-dimensional graviton. The RSII case is
easier as there the fermionic higher spins are chiral, in particular,
imposing the condition $\Psi_{5...}=0$, \cite{GK}
\be
\Psi^R_{a_1...a_{s-1/2}}=0\ ;\ \Psi^L_{a_1...a_{s-1/2}}=\sqrt{4a(s-1/2)}e^{-s\sigma}\psi^L_{{a_1...a_{s-1/2}}}\ ,
\ee
where
\be
\tilde{\gamma}^b\partial_b \psi_{a_1...a_{s-1/2}}=0\ ,
\ee
and $\tilde{\gamma}^b$ are the Dirac matrices in Minkowski. For the graviton
we have, similarly imposing $h_{5\mu}=0$,
\be
h_{ab}=2a e^{-2\sigma}\zeta_{ab}\ ,
\ee
where
\be
\square \zeta_{ab}=0\ ,
\ee
and the bulk metric is $g_{ab}=g_{ab}^{\tiny AdS}+2M_5^{-3/2}
h_{ab}$. The five dimensional Planck mass is $M_5$ and $\square$ is
here calculated in Minkowski.

The chirality of the HS fields implies that the only non-zero terms in
the action (\ref{fermionic}) are the ones containing an odd number of
gamma matrices.
By dimensionally reducing the action (\ref{fermionic}) and imposing gauge conditions,
we therefore obtain
\be
S_{\mbox{\tiny reduced}}=\frac{s}{2}M_5^{-3/2}\sqrt{2a}\int d^4x \sqrt{-q}\bar\psi_{a_1...a_{s-1/2}}\tilde{\gamma}^a\eta^{a_1 c}
\left[\zeta_{b a,c}+\zeta_{b c,a}-\zeta_{ac,b}\right]\psi^{ba_2..a_{s-1/2}}\ .
\ee
Using the RS fine tuning $12G_N=M_5^{-3}a$ where $G_N$ is the four-dimensional Newtonian constant, we easily find that
\be
S_{\mbox{\tiny reduced}}=s \sqrt{24\pi G_N}\int d^4x\sqrt{-q} \bar\psi_{a_1...a_{s-1/2}}\tilde{\gamma}^a\eta^{a_1 c}
\left[\zeta_{b a,c}+\zeta_{b c,a}-\zeta_{ac,b}\right]\psi^{b a_2..a_{s-1/2}}\ .\label{sca}
\ee
The scattering amplitude
obtained from (\ref{sca}) is therefore suppressed at the Planck scale. The massive KK cases are more complicated, however by dimensional analysis
one can infer that the scattering amplitudes are still suppressed by the Planck scale. The volume of the infinite tower of continuum KK
modes is not enough to counteract the Planck scale suppression. In
fact, the density of the massive KK modes is exponentially suppressed
as it is in the
graviton case \cite{Randall:1999ee}.

Therefore, if HS unparticles are used as Dark Matter candidates, a
mechanism to obtain the observed abundances must be found. Although
the model presented here captures
many features of a more general stringy model, it is over-simplified. In
particular, the only bulk matter considered here is a cosmological constant. In String
theory however other bulk fields may couple with the HS. These
interactions (possibly inflaton decay) may produce the density of HS
particles necessary for
HS to be plausible DM candidates.
We leave this for future research.

\section{Conclusions}

Massless higher spin fields are generically inconsistent in a curved background. However Vasiliev \cite{Vass}
proved that by an infinite
series of non-minimal interactions of HS and curvature tensors on a constant curvature background,
gauge invariance may be restored. However,
the extraction of single terms of the expansion from the formalism of \cite{Vass} is an extremely difficult task.
In this paper we have therefore followed a different path.

Correcting and generalising the work of \cite{GK},
we constructed consistent interactions of gravity and branes with bosonic and fermionic higher spin fields by
a mechanism similar to the one proposed by \cite{Cucchieri:1994tx} for
massive HS interacting with gravitons in a flat background.
In \cite{Cucchieri:1994tx} an expansion of non-minimal terms coupling HS ($\phi$) with gravity in the dimensionless parameter
$\phi {\cal R}\phi/m$ was made, where ${\cal R}$ is a generic tensor depending linearly on
the curvature of the spacetime and $m$ is the gauge-breaking mass. In the $AdS$ case, HS fields are gauge-invariant.
However, a non-derivative term proportional
to the cosmological constant $\Lambda$ in the free HS action naturally appears.
One can therefore, following the main idea of \cite{Cucchieri:1994tx},
expand the HS interaction with gravity in powers of $\phi {\cal R}\phi/\sqrt{\Lambda}$.
With this in mind, we found consistent tree-level interactions of higher spin fields with gravity on an $AdS$ background
preserving the unitarity bound of the graviton-HS Compton scattering amplitudes
up to the Planck scale. A full restoration of the free HS gauge invariance
when interacting with gravity, although theoretically important,
is left for future work. Nevertheless, if only phenomenological purposes are in mind,
the lagrangians constructed here well describe the interactions of HS
fields
with linearised gravity up to the Planck scale.

In the braneworld case, in which the bulk is bounded by
a brane where the Standard Model lives, we showed that an
on-shell linearised
gauge-invariant action for fermionic higher spin fields can be found. In the bosonic case a gauge-invariant
action cannot be
constructed unless the brane is a pure-tension brane or if Dirichlet
boundary conditions are imposed on the HS field on the brane.
We concluded that a brane observer can only measure fermionic higher spin unparticles \cite{giorgi}.
The unparticle behaviour of
these fields, as observed by a brane observer, comes from the fact
that the KK decomposition of HS consists of a massless mode and a continuum of
massive modes, without a mass gap.

In the last part of our work we showed that fermionic higher spin
fields cannot interact with brane SM, only with brane gravitons.
We considered their decay into (production by) brane gravitons. We showed that
HS may be produced by graviton scattering only at Planckian
scales. This fact makes
HS fields stable during the classical evolution of the Universe in braneworlds.
This stability might promote HS as possible Dark Matter candidates.
However a mechanism for producing the observed abundances of Dark Matter out of HS must be found.
We leave this for future research.

\section*{Note Added}
While we were preparing this work, we were made aware of
\cite{foto}, which provides a more general approach to calculate HS
interaction vertices. Their approach has been applied to scalars
interacting with HS \cite{foto2} and it would be interesting to see
how our results compare to their approach applied explicitly to the
graviton vertex. However this is beyond the scope of this paper.

\acknowledgements
CG wishes to thank Alex Kehagias for useful comments on the extension
of the work of \cite{GK}. CG wishes to thank Marco Serone for discussions on unparticles scattering processes.
CG also wishes to thank Torsten Bringmann for discussions on Dark
Matter candidates. Finally CG wishes to thank King's College, London,
Physics Department for hospitality during part of this work.
CG and AS wish to thank Ben Allanach and Tom Varley for many useful discussions on basic aspects of particle physics.
AS wishes to thank Malcolm Perry for reading an early draft of the
manuscript. AS also wishes to thank PPARC for support during part
of this work.

\appendix

\section{Non-minimal actions}

In this appendix we give a brief outline of the calculation of the
non-minimal linearised
on-shell
gauge-invariant fermionic and bosonic actions in an $AdS_5$ background.

To find the actions, we calculated the gauge variation of the minimal actions ((\ref{as}) and (\ref{zero})) under (\ref{generalvar})
and (\ref{df}) respectively on a perturbed $AdS$ background. As already stressed before, this variation has residual non-vanishing terms that
have to be cancelled by appropriate non-minimal counter-terms.

Let us firstly consider the bosonic higher spin fields. We only consider tree-level gauge invariance, {\it i.e.} we consider only
gauge invariance when both the HS fields and gravitons (perturbations
away from AdS) are on-shell. Once we have imposed the gauge constraint $
\Phi^{\mu}\!_{\mu\alpha_3\ldots\alpha_s}=0$, the free equations of motion for the graviton and
the HS fields are:
\be
R_{\mu\nu}=-\Lambda g_{\mu\nu},
\label{heom}
\ee
\be
(\nabla^2-M^2)\Phi_{\alpha_1\ldots\alpha_s}=0
\ee
and
\be
\nabla^{\mu}\Phi_{\mu\alpha_2\ldots\alpha_s}=0.
\ee

In addition to these we use standard symmetries of the Riemann
tensor. The most useful of these for spacetimes obeying (\ref{heom})
is
\be
\nabla_{\mu}R^{\mu\nu\rho\sigma}=0.
\ee
Then the gauge variation of the minimal action, (\ref{zero}), under
(\ref{df}), after the free equations of motion and gauge constraints have been used, is
\be
\delta S_0^b = -2(s-1)\int d^5\!x \xi_{\alpha_2\ldots}R_{\alpha_1}\!^{\mu\alpha_2}\!_{\tau}\nabla_{\mu}\Phi^{\alpha_1\tau\alpha_3\ldots}\ .\label{A5}
\ee
The non-minimal term necessary to cancel (\ref{A5}) can be guessed by
generalizing \cite{Cucchieri:1994tx} to our case. To modify their calculation for 5d $AdS$, we took their non-minimal
action and replaced Riemann tensors by Weyl tensors, so that the
non-minimal terms vanish for the unperturbed background
spacetime. This change is all that is necessary. Equations (\ref{snm})
and (\ref{DSB}) in the main text give the modified non-minimal terms,
$S^b_{nm}$ and $\Delta S^b$. With these two terms added to the minimal
action, the variation of the total action, $S^b=S^b_0+S^b_{nm}+\Delta S^b$, is

\be
 \delta S^b = \frac{(s-1)(s-2)}{8a^2(s+1)}\int d^5x \xi_{\alpha\gamma}\!^{\mu_4\ldots}(\nabla^2-M^2)
 ((\nabla^{\rho}W^{\alpha\beta\gamma\delta})\Phi_{\beta\delta\rho\mu_4\ldots}).
\ee
This is proportional to an equation of motion and so can be removed by a local field redefinition of $\Phi$ (as in \cite{Cucchieri:1994tx}).

Let us now consider the fermionic higher spin fields. In repeating
\cite{Cucchieri:1994tx}'s calculation for flat 4d spacetime we found
their non-minimal action can be simplified. Instead of adding
\be
S_{nm} = \frac{n(n-1)}{2m}\int d^4x
\bar{\Psi}^{(n)\alpha\gamma\mu_3\ldots\mu_n}R^+_{\alpha\beta\gamma\delta}\Psi^{(n)\beta\delta}\!_{\mu_3\ldots\mu_n}
\label{CDPSnm}
\ee
and
\be
\Delta S= \frac{2n(n^2-1)(n-2)}{m^3(2n+1)}\int d^4x
\bar{\Psi}^{(n-1)\alpha\gamma\mu_4\ldots\mu_n}\left(\partial^{\lambda}(\displaystyle{\not}\partial+m)
R^-_{\alpha\beta\gamma\delta}\right)\Psi^{(n)\beta\delta}\!_{\lambda\mu_4\ldots\mu_n},
\ee
where $\Psi^{(n)}_{\mu_1\ldots\mu_n}$ is  the spin $s=n+\frac{1}{2}$
field, $\Psi^{(i)}_{\mu_1\ldots\mu_i}$ $i<n$ are the auxiliary fields
necessary for the description of massive higher spin fields, but not necessary in
our massless case, and
$R^{\pm}_{\mu\rho\nu\sigma}=R_{\mu\rho\nu\sigma}\pm
\frac{1}{2}\gamma^5\epsilon^{\nu\sigma\alpha\beta}R_{\alpha\beta}\!^{\mu\rho}$,
to obtain a non-minimal action, $S_{nm}$,(\ref{CDPSnm}), and
\be
\Delta S= \frac{2n(n^2-1)(n-2)}{m^2(2n+1)}\int d^4x
\bar{\Psi}^{(n-1)\alpha\gamma\mu_4\ldots\mu_n}\left(\partial^{\lambda}R^+_{\alpha\beta\gamma\delta}\right)
\Psi^{(n)\beta\delta}\!_{\lambda\mu_4\ldots\mu_n},
\label{CDPDSimproved}
\ee
work equally well.

Naively we might now think that we could proceed as we did for the
bosonic fields and just replace the Riemann
tensors by Weyl tensors in (\ref{CDPSnm}) and (\ref{CDPDSimproved}) to
obtain a non-minimal action for 5d AdS. This doesn't work because the
identity
$W_{\alpha\beta\rho}\!^{\left[\mu\right.}\gamma^{\left.\nu\right]\sigma\alpha\beta}=0$
no longer holds in 5d. Instead one needs a more complicated action
which is given in
(\ref{sf}) and (\ref{DSF}). This corrects the action of \cite{GK}.

To calculate the gauge variation of the total action,
$S^f=S^f_0+S^f_{nm}+\Delta S^f$, we use the free equation of motion of
$\Psi$:
\be
\gamma^{\rho}D_{\rho}\Psi_{\alpha_1\ldots\alpha_{s-\frac{1}{2}}}+2as\Psi_{\alpha_1\ldots\alpha_{s-\frac{1}{2}}}=0.
\ee
where the restricted gauge constraint on $\Psi$ has been imposed.

Using these, the graviton equation of motion and identities of the
Riemann tensor,
\be
\delta S^f = \frac{\left(s-\frac{3}{2}\right)\left(s-\frac{5}{2}\right)}{8a^2\left(s+\frac{1}{2}\right)}\int d^5x
\bar{\epsilon}_{\mu\nu\alpha_4\ldots}(\bar{\displaystyle{\not}\nabla}+2as-5a)
((\bar{\nabla}_{\alpha_3}{\cal W}^{\mu\rho\nu\sigma})\Psi_{\sigma\rho}\!^{\alpha_3\ldots}),
\ee
which can again be eliminated by a local field redefinition of $\Psi$ (as in \cite{Cucchieri:1994tx}).

\end{document}